# Magneto-optical imaging of stepwise magnetic domain disintegration at characteristic temperatures in EuB$_6$

Dibya J. Sivananda[1], Ankit Kumar[1], Md. Arif Ali[1], S. S. Banerjee[1,*]

[1]*Department of Physics, Indian Institute of Technology Kanpur, Kanpur 208016, India.*

Pintu Das[2]

[2]*Department of Physics, Indian Institute of Technology Delhi, New Delhi 110016, India.*

Jens Müller[3]

[3]*Institute of Physics, Goethe-University Frankfurt, 60438 Frankfurt (M), Germany.*

Zachary Fisk[4]

[4]*Department of Physics, University of California, Irvine, California 92697, USA.*

*email: satyajit@iitk.ac.in

*Abstract:*

Prior to the onset of the ferromagnetic transition in semimetallic EuB$_6$, unusual magnetic and electric behavior have been reported. Using a highly sensitive magneto-optical imaging (MOI) technique, we visualize the behavior of magnetic domains in a EuB$_6$ single crystal. The transformation from a paramagnetic to a ferromagnetic state is shown to be non-Curie Weiss like and proceeds via multiple breaks in the curvature of the temperature-dependent local magnetization. From our experiments, we identify three characteristic boundaries, $T^*(H)$, $T^*_{c1}(H)$ and $T_{c2}(H)$, in a field - temperature magnetic phase diagram. Using scaling and modified Arrott's plot analysis of isothermal bulk magnetization data, we determine the critical exponents $\beta$ =

0.22±0.01, $\gamma = 0.88±0.05$ and $\delta = 5.0 ±0.1$ and a critical transition temperature = 12.0 ± 0.2 K which is found to be equal to $T_{c2}$. The critical exponents are close to those associated with the universality class of tricritical mean field model. The absence of a model with which the exponents correspond with directly, suggests the presence of large critical fluctuations in this system. The critical fluctuations in this system are sensitive to the applied magnetic field, which leads to field dependence of boundaries in the magnetic phase diagram. Deep inside the ferromagnetic state at T below $T_{c2}$, we observe the presence of large magnetized domains along with the observation of Barkhausen like jumps in local magnetization. With increasing T the magnetic domains disintegrate into finger-like patterns before fragmenting into disjoint magnetized puddles at $T_{c1}^*$ and ultimately disappearing at $T^*$. At $T_{c1}^*$ we observe a significant increase in the spatial inhomogeneity of the local magnetic field distribution associated with the magnetic domain structure disintegrating into smaller magnetized structures. We explain our results via the formation of magnetic polaronic clusters and their coalescing into larger domains.

*Introduction:*

In systems with strong electron correlations, there are transformations from a disordered to an ordered spin configuration. The correlations impact not only their magnetic but also electrical properties significantly. A case in point are rare-earth hexaborides ($RB_6$, R = Ca, Sr, Ba, La, Ce, Sm, Eu, and Gd). These are correlated electron systems which display diverse magnetic and electrical properties [1,2]. This class of materials exhibits strong electron-electron correlations due to narrow, incompletely filled *d* or *f* bands. The properties of these materials are quite diverse, ranging from superconductivity found in $YB_6$ [3] to complex antiferromagnetic phases with Kondo like features in CeB$_6$ [4, 5] and low carrier density ferromagnetism in semi-metallic EuB$_6$ [6], as well as an exotic Kondo-like topological insulating state reported in SmB$_6$ [7,8,9]. Amongst these materials, in EuB$_6$ a complex exchange interaction exist between the itinerant *d*-electrons and localized *f*-electron moments of Eu$^{2+}$, producing a charge localized state with significant local magnetization, viz., magnetic polarons. The formation and size of a magnetic polaron is governed by a balance between kinetic energy of the itinerant *d* - electrons and exchange interaction between the itinerant electrons and the localized *f* - electron moments of Eu [10]. Traditionally the concept of magnetic polarons was introduced to study metal-insulator transition in EuO [11]. EuB$_6$ exhibits a transformation from a paramagnet at higher temperatures to a ferromagnetic semi-metal at lower temperatures [6]. Electrical transport, specific heat and optical measurements in EuB$_6$ have shown, anomalies near $T_{c1}$~ 16 K and $T_{c2}$ ~ 12 K [12,13,14,15,16,17], where below $T_{c2}$ the system is ferromagnetically ordered. At these temperatures, one finds large magnetoresistance, enhanced electronic noise, non-linear electrical transport properties [18], development of peaks in the Raman spectra [19], and a pronounced lattice distortion [20]. These observations have led to suggestions of percolation of magnetic polarons throughout the system [18,21,22,23,24,25,26,27,28,29,30,31] at $T_{c1}$. While EuB$_6$ has been extensively characterized with bulk magnetization and transport measurements, few studies exist on imaging the local magnetic properties in this system to study the possible effects of magnetic polaron formation on the behavior of magnetic domains in this system. Using high sensitivity Magneto-Optic Imaging (MOI) technique, we map the local magnetic field distribution in a high-quality EuB$_6$ single crystal. Analysis of the temperature dependence of the local magnetic field shows significant non - Curie-Weiss behavior with breaks in curvature at characteristic temperatures. In a field (*H*) – temperature (*T*) phase diagram, we identify three boundaries $T_{c2}(H)$, $T_{c1}^*(H)$ and $T^*(H)$. At low $T$, below $T_{c2}$ the sample is in a ferromagnetic state with large sized domains and is in a paramagnetic state at high $T$, above $T^*$. We investigate the behavior of fluctuations in this system using critical scaling and modified Arrott's plot analysis of isothermal bulk magnetization data. Using these we determine critical exponents $\beta = 0.22\pm0.01$, $\gamma = 0.88\pm0.05$ and $\delta = 5.0 \pm 0.1$ and a critical transition temperature $= 12.0 \pm 0.2$ K which is found to be equal to $T_{c2}$. The critical exponents are close to that associated with the universality class of tricritical mean field model, however, the observed exponent values are not exactly equal to the latter. The absence of a direct match with a mean field model suggests the presence of large critical fluctuations in this system. We observe large magnetic domains below $T_{c2}$ with the local magnetization exhibiting Barkhausen like jumps. Between $T_{c2}$ and $T_{c1}^*$ we observe the magnetized domains fragmenting into finger-like patterns. At $T_{c1}^*$ the domains disintegrate further upon increasing temperature into disjoint magnetized puddles with their density becoming undetectable above $T^*$. We show that the temperature window over which these isolated magnetized puddles exist increases with $H$. At $T_{c1}^*$ we observe a significant increase in the spatial inhomogeneity of the local magnetic field distribution. Our results are explained on the basis of formation of clusters of magnetic polarons which subsequently coalesce to form larger domains. In the phase diagram close to the critical transition regime $T_{c2}(H)$ we identify the $T_{c1}^*$ ($H$) boundary where coalescence of polaron clusters happen. We show that $T_{c1}^*(H)$ is distinct from another $T_{c1}(H)$ boundary, which is where disjoint individual polarons reach a sizeable dimension upon lowering of the temperature from $T^*$. Our analysis also suggests that applied magnetic fields affect critical fluctuations in the system leading to the observed features in the phase diagram.

*Experimental Results and Discussions:*

High-quality EuB$_6$ single crystal with dimensions $1.0 \times 0.5 \times 0.1$ mm$^3$ were grown using Al flux method [26]. Cleaved samples from the same batch of single crystals have been used for a variety of specific heat, bulk magnetization and nonlinear transport and resistance noise studies [18,20,26] (see [32] supplementary information section-I). Using electron probe microanalysis (EPMA) we find the sample is chemically homogeneous. Reference [32] supplementary information section-II shows EPMA concentration fluctuations of Eu measured across 12 different regions distributed around the sample (see locations marked 1 to 5) is constant to within a standard deviation of 0.5%. Using a highly sensitive MOI technique [33], we map out the local magnetic field distribution $B_z(x,y)$ across the crystal surface at different applied magnetic fields $H$ ( || [001]) and temperatures $T$ (where $z$-axis is along the [001] direction and perpendicular to the crystal's flat surface ($x,y$)). Briefly, in this technique, a linearly polarized beam of light is reflected off the surface of a freshly cleaved surface of EuB$_6$ on which a Faraday active film is placed. The film experiences the local magnetic field $B_z(x,y)$ from the sample which Faraday rotates the light. The intensity distribution of the reflected, Faraday rotated light, $I(x\ y)$, is captured using a combination of polarizers and high sensitivity CCD camera [33] (for more details see [32] supplementary information section III). From $I(x\ y)$ the local magnetic field distribution is determined (note $I(x,y) \propto [B_z(x,y)]^2$, within the field range of our investigations) using suitable calibration,

Figure 1(a) shows the bulk magnetization behavior of the investigated EuB$_6$ crystal measured in a SQUID magnetometer (Cryogenics, UK). The magnetization response of this sample is comparable to that reported earlier [23]. Inset of fig.1(a) shows the magnetization versus temperature, measured at $H$ = 10 Oe. The bulk $M(T)$ behavior, is Curie-Weiss like with a paramagnetic Curie temperature $T_\theta$ = 13.7 ± 0.4 K, below which EuB$_6$ exhibits bulk ferromagnetic (FM) ordering. In the main panel of fig.1(a) we show $M(H)$ measured at 9K, 13K and 17K. At 9 K the $M(H)$ has an almost square shape. The hysteresis in $M(H)$ is weak and the coercive field at 9 K was estimated to be below 50±10 Oe, where the error is due to the remnant field of 10 Oe in the superconducting magnet of the SQUID magnetometer. At higher $T$, the $M(H)$ loop's shape begins to get skewed. Figure 1(b) shows the raw MOI of the sample at 9.1 K (in the FM state). All our MOI are captured in the forward leg of the magnetization curve, viz., after reaching 9.1 K, $H$ = -600 Oe is applied with MOI captured as $H$ is increased towards +600 Oe (600 Oe is close to the saturation field of EuB$_6$). For better visualization, the region outside the sample have been intentionally darkened. The bright to dark contrast variations visible inside the sample represent changes in the intensity of Faraday-rotated light, viz., a higher intensity (or brighter contrast) represents larger local $B_z$ and vice versa. The measured $B_z(x, y)$ distribution across the sample is color coded in fig. 1(c). Figure 1(c) shows the presence of large domains with uniform magnetization across the sample. In the MOI of fig.1(c) taken deep in the FM state of the sample, the domains with yellow contrast (larger $B_z$) are domains magnetized nearly perpendicular to the sample surface (H ∥ [001] direction of the sample). The bluish shade regions have magnetization oriented away from [001] direction, resulting in a relatively smaller $B_z$ over these regions.

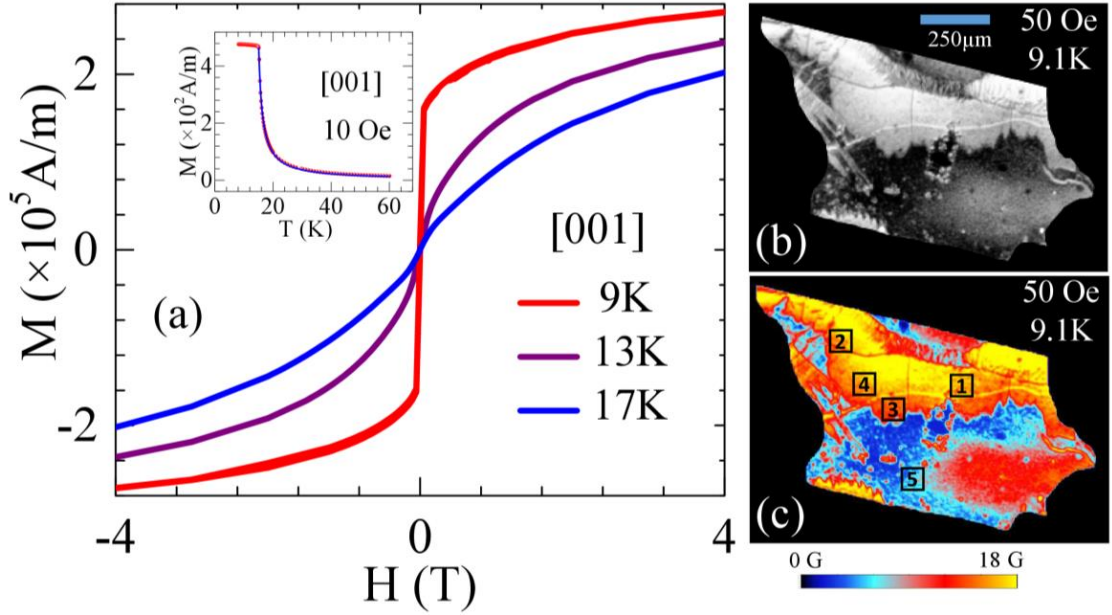

Figure 1(a): M-H curves at 9K, 13K, and 17K. The inset shows magnetization with temperature. (b) An MOI at 9.1K and 50 Oe. (c) Colour coded image of (b). Locations 1 ($A_1$), 2 ($A_2$), 3 ($A_3$), 4 ($A_4$) and 5 ($A_5$) mark positions where $<B_z>$ is determined. The sample locations with higher $B_z$ values are colored yellow, intermediate $B_z$ values are red, and low $B_z$ values dark blue.

Figures 2(a) to 2(d) show that with increasing $H$ at 9.2K, the area covered by bright yellow and reddish regions (i.e., sites with stronger $B_z$) spreads across the entire sample by 150 Oe, at the expense of bluish regions, (viz., domains with magnetization oriented perpendicular to the sample begin spreading across the sample as $H$ is increased). Figures 2(e) and 2(f) show that at a constant $H = 50$ Oe, as the $T$ increases from 9.1K to 13K, some of the magnetized regions with relatively large $B_z$ (bright yellow and reddish sites) begin expanding into finger-like structures (see regions inside the dashed rectangle and also region near the lower sample edge). Figure 2(i) shows the overlapped image of the image portions present within the dashed rectangle of figs.2(e) and (f). The region between the black and the green curves in fig.2(i), represents the spread of the domain upon reaching 13K from 9.1K. With further increase in $T$ (fig.2(g)) the $B_z$ over the domains weakens, and the domains begin to shrink rather than expand (see fig.2(h)). The domain shrinkage is seen inside the dark dashed circle in figs.2(g) and 2(h). At higher $T$ the domains gradually disappear from the image. We would like to mention that the images in fig.2 show the presence of a horizontal fault line spreading across the sample. Our results in fig.2 show no correlation between the growth and propagation of the magnetic domains (clusters) across the sample with the location of these elongated defects extending across the sample [see also figs 1(b) and 1(c)]. In fact, the directions in which the magnetic domains propagate out and spread out with changing magnetic field or temperature, have no correlation with the direction of the defect. This also helps us to show that local physical inhomogeneities play no role in generating the domain features discussed in this paper. Furthermore, fig.2(d) shows that by 150 Oe, the average local $B_z$ becomes almost uniform across the entire sample, thereby supporting our assertion that the sample is chemically homogeneous, and that the bottom half of the sample shown in the MOI image of fig.2, doesn't behave any differently from the top half.

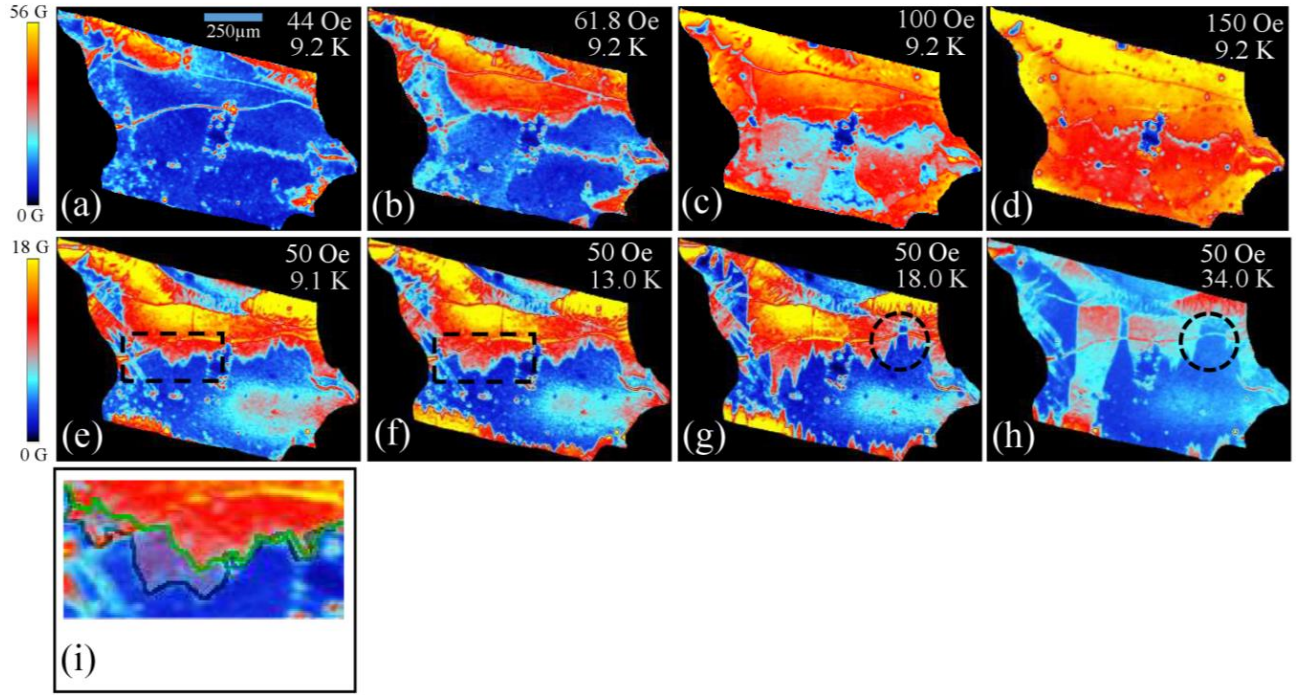

Figure 2: (a)-(d) Colour coded MO images at constant 9.2K and varying field. (e)-(h) Colour coded MO images at constant 50 Oe and varying temperature. Dashed rectangle and circle show the portions of the domain where their growth or shrinking is visible. (i) Shows region inside the dashed rectangle in (e) and (f) overlapped over each other with 50% transparency. Green curve indicates the contour of the magnetic domain edge at 9.1K (see (e) contour of the red region inside the dashed rectangle), and the black curve indicates contour of the magnetic domain at 13K (see (f) contour of the red region inside the dashed rectangle).

In fig.3 we study the temperature dependence of the local magnetization in the sample by measuring the average local magnetic field, i.e.,

$$<B_z>(T) = \frac{\int B_z(x, y) dx dy}{\int dx dy},$$

where the area of integration is a 50μm×50μm square region, located at different regions of the sample, see locations 1 to 5 in fig.1(c)). Figures 3(a) and (b) compare the $<B_z>(T)$ behavior measured at 50 Oe at different locations $A_1$ and $A_2$ on the sample (see the location of $A_1$ to $A_5$ in fig.1(c)).

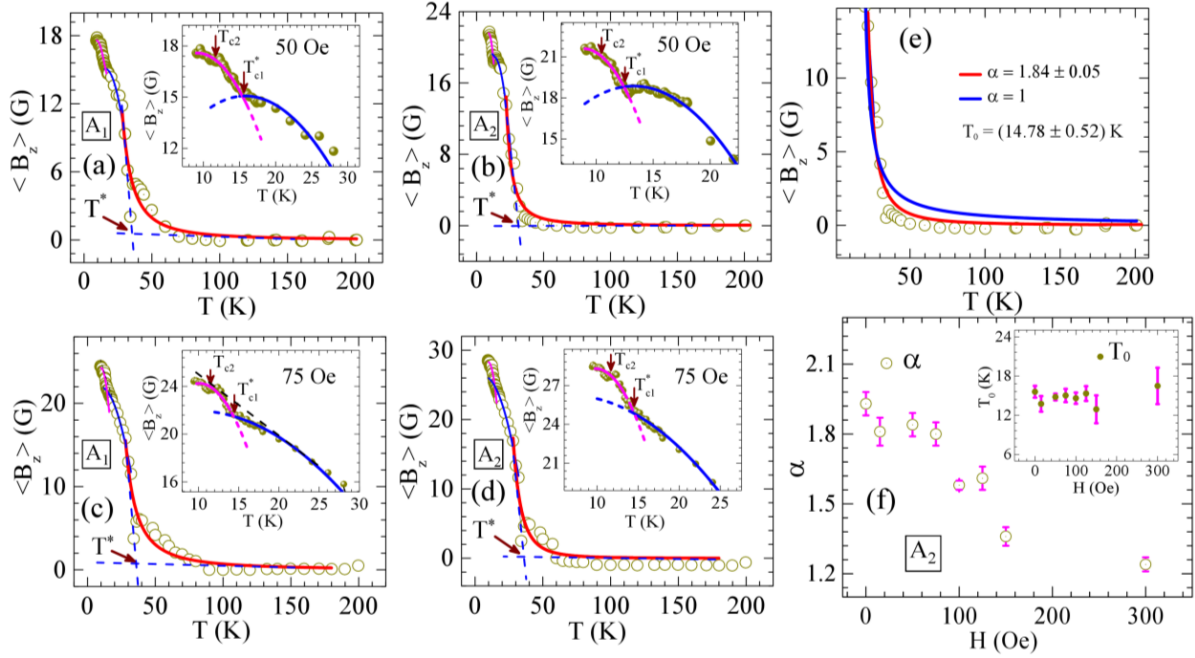

Figure 3: (a)-(d) $<B_z>$ with temperature, for location 1 ($A_1$) and 2 ($A_2$); Insets in (a)-(d) show the part of the corresponding main panel below 30K with $T_{c1}^*$ and $T_{c2}$ identified by brown arrows. $T^*$ is identified in the main panel (a)-(d) with brown arrow. (e) Shows the data in (b) being fitted by a non-Curie-Weiss (red) and Curie-Weiss (blue) curve. (f) Fit parameters to eq. (1): Variation of α and $T_0$ (inset) with field.

From fig.3 note that, $<B_z>$ increases rapidly below 70 K for region $A_1$ and below 50 K for region $A_2$ (see figs. 3(a) and 3(b)). In the presence of higher applied magnetic field, of 75 Oe, the $<B_z>(T)$ starts to increase below 100K for $A_1$ and below 70K for $A_2$ (see figs.3(c) and 3(d)). A comparison of the 50 Oe and 75 Oe data shows that the increase in $<B_z>(T)$ is more gradual at higher $H$ (we have observed this at higher fields also, data not shown). The $<B_z>(T)$ in figs. 3(a) – (d) fits (see red curve in fig.3(a)-(d)) to a form which suggests critical behavior:

$$\langle B_z \rangle (T) = \frac{C}{|T-T_0|^\alpha}, \qquad (1)$$

where $C$ is a constant, $T_0$ is a characteristic temperature scale ($\neq T^*$) and $\alpha$ is a fitting exponent. The data in figs.3(a) - (d) best fits eqn. (1) with $\alpha \sim 2$ and $T_0 \sim 15$ K. In fig.3(e) we compare the $<B_z>(T)$ data of fig.3(b) fitted with $\alpha = 1.84$ in eq. (1) (red curve) and a Curie Weiss form (see blue curve), i.e. $\alpha = 1$ (for both fittings we use $T_0 = 14.78 \pm 0.52$ K). Figure 3(e) clearly shows that the best fit to the local magnetic fields, $<B_z>(T)$, is non – Curie-Weiss. Figure 3(f) shows that, $T_0 \sim 14$ to 15 K is nearly independent of $H$, while $\alpha$ is sensitive to $H$. At low $H$ the value of $\alpha \sim 2$, whereas above 100 Oe $\alpha$ decreases and approaches 1, which is consistent with our observation that $<B_z>(T)$ increases more gradually at higher fields. This feature suggests that the critical behavior in the system is affected by magnetic fields (we will explore this feature later). The transition $T_\theta$ determined from fig.1 is only a gross estimate, and we will improve upon this estimate later. The local magnetization versus temperature data has a number of features, which we study by defining a few characteristic temperatures. In fig.3, we identify another temperature, $T^*$, as the intersection of the low-$T$ and high-$T$ extrapolations of the $<B_z>(T)$ behavior.

Upon lowering the temperature, at $T^*$ the magnetization increases due to the onset of enhanced magnetic correlations in the sample prior to the transformation into a ferromagnetic state below $T_0$ (or $T_\theta$). Insets of figures 3(a)-(d), show that below 30 K the local $<B_z>(T)$ has abrupt breaks in curvature at two locations marked as $T_{c1}^*$ and $T_{c2}$ (see the break in the curvature of the solid lines drawn through the data points in the insets of fig.3). Note, the solid lines are a guide to the eye, and are empirical second order polynomial fits to $<B_z>(T)$ data in the different $T$ regimes. Below the inflection, in the $<B_z>(T)$ curve marked as $T_{c2}$ the magnetization saturates indicating the onset of bulk FM order in the sample. The insets also show the presence of another significant inflection at $T_{c1}^* \sim 15$ K. Our $T_{c2}$ and $T_{c1}^*$ determined by the criteria of inflection in the $<B_z>(T)$ curve compare well with the values of these temperatures obtained from bulk measurements by others [12-27].

Figures 4(a) and 4(b) show the field dependence of $T_{c1}^*$, $T_{c2}$ and $T^*$ for $A_1$ and $A_2$ regions (their locations are shown in fig.1(c)). We observe that $T_{c1}^*$ and $T_{c2}$ decrease almost linearly while $T^*$ increases with $H$. We note that across different regions of the sample, the values and the field dependence of $T_{c1}^*$, $T_{c2}$, and $T^*$ are similar, indicating a homogenous sample. We find similar phase diagrams as fig.4 across three other different locations ($A_3$ to $A_5$, see their location in fig.1(c)) on the sample (see [32] supplementary information section-IV for the phase diagrams). By comparing these phase diagrams in regions $A_1$-$A_5$ (which are identical to the regions where EPMA measurements were carried out, see [32] supplementary information section-II), we show that at a given field the values of $T^*$, $T_{c1}^*$ and $T_{c2}$ are nearly identical across different locations on the sample. This also establishes the uniformity of the behavior across the sample and the homogenous quality of the sample.

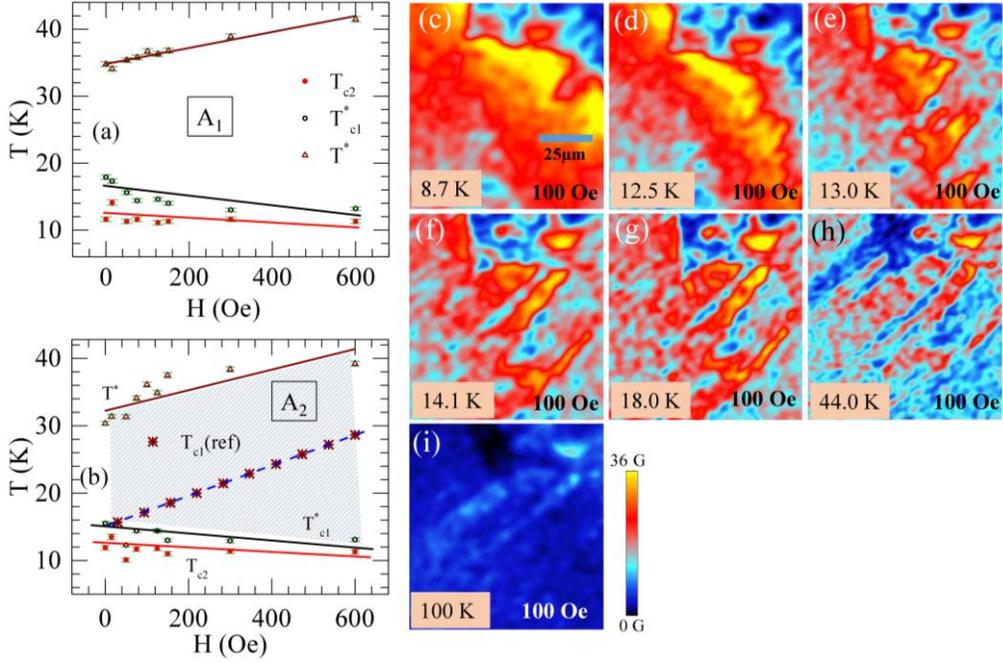

Figure 4:(a) and (b) The phase diagrams at location 1 ($A_1$) and 2 ($A_2$). The lines drawn through the data points (red, black and brown lines) which are labeled as $T_{c2}$, $T_{c1}^*$ and $T^*$ are a guide to the eye. Dashed (blue) line in (b) represent $T_{c1}$ line as determined in reference [18]; (c)-(i) High-resolution MO images with varying temperature. The site of these images is at the center of the sample just below location 1 in fig.1 (c).

High sensitivity high-resolution MO images (of size 100μm×100μm) in fig.4(c)-(i) show the temperature dependence of the relatively strongly magnetized regions in the sample. Compared to fig.2, these images are captured with a higher optical magnification at 100 Oe (from a region near the center of the sample). The images show that from 8.7 K to $T_{c2}$ = 12.5K the yellow and red colored magnetic domains do not change appreciably. Below $T_{c2}$ it is known from earlier studies as well as our studies (fig.1), that EuB$_6$ is in an FM state (we observe Barkhausen like jumps in local $<B_z>$ below $T_{c2}$, see fig.5(c)-(d)). The large domains below $T_{c2}$, fragment into smaller finger-like structures at $T > T_{c2}$ = 12.5 K. With further increase in $T$ beyond $T_{c1}^*$ ~ 15 K (see figs.4(g) and 4(h)) one observes the finger-like structures dissociate into smaller distinct puddle-like magnetized entities. The puddles have larger local $B_z$ (yellowish or reddish puddles) surrounded by light red or bluish regions which have relatively smaller $B_z$ (interestingly, fig.5(a)-(b) show an enhanced scatter in the local $B_z$ values at $T_{c1}^*$ as the domains begin fragmenting into puddles). With increasing $T$, the puddles become smaller and begin disappearing with the local $B_z$ within the puddles also decreasing. At 44 K (fig.4(h)) there are very few yellowish puddles one can identify. At $T >> T^*$, viz., at 100 K (fig.4(i)) the $B_z$ across the sample uniformly approaches zero.

Figures 5(a), (b) show the behavior of the mean square deviation ($\delta^2$) in the $B_z$ distribution as a function of temperature with $\delta^2(T) = (<B_z>(T)$ – average trend of $<B_z>(T))^2$. Note the average trend of $<B_z>(T)$ value is the best fit line in figs. 3(a) - 3(d) insets. In these figures, the location of $T_{c2}$ and $T_{c1}^*$ (as determined from the phase diagrams of fig.4(a) and (b)) have been marked. Figures 5(a) and (b) show a significant enhancement in the scatter of the average local field values, $\delta^2(T)$ near $T_{c1}^*$. Below $T_{c2}$ and in between $T_{c2}$ and $T_{c1}^*$, the $\delta^2(T)$ is significantly lower. Near $T_{c1}^*$ the enhanced $\delta^2$, is related to the enhancement in the scatter of $B_z$ values as relatively large uniformly magnetized domains dissociate into individual puddles and weaken in strength. The changing statistics of these fragmented domains with temperature leads to an enhanced spatial inhomogeneity in the $B_z$ distribution which leads to enhanced $\delta^2$ near $T_{c1}^*$. Apart from using the criteria of change in curvature of $<B_z>(T)$ curves in fig.3, the enhancement in $\delta^2(T)$ is also used to determine the locations of $T_{c2}$ at different $H$ (see [32] section-V of supplementary information). The two criteria for determining $T_{c2}$ are found to be mutually consistent (the weakly field

dependent $T_{c2}$ is also determined from the analysis in fig.6, and the different ways of determining $T_{c2}$ are consistent)

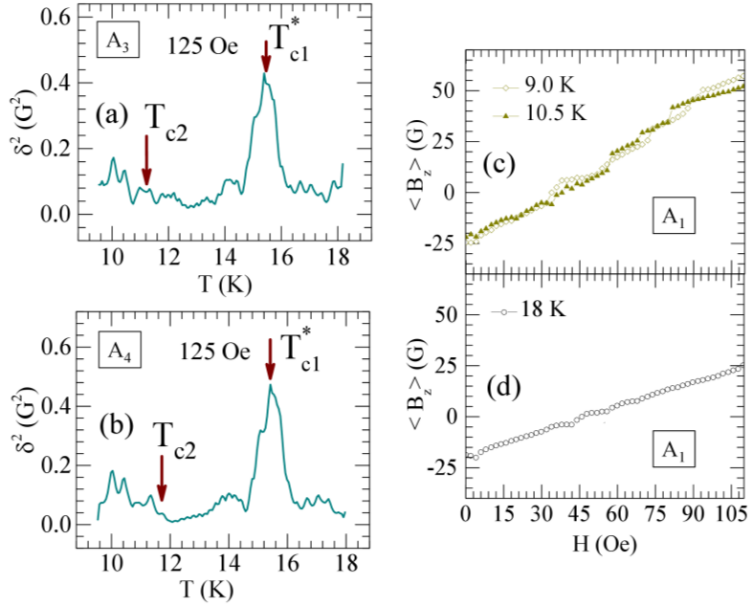

Figure 5: (a)-(b) The mean square deviation ($\delta^2$) versus temperature curves show a sudden increase of the inhomogeneity in the local $<B_z>$ values as one crosses cross $T_{c1}^*$ and $T_{c2}$. The positions of $T_{c1}^*$ and $T_{c2}$ as determined from magneto-optics is indicated by brown arrows; (c) and (d) $<B_z>$ versus external field showing Barkhausen jumps at 9K and 10.5K and no jumps at 18K.

The $<B_z>$ versus $H$ determined from MOI in figs.5(c),(d), shows the presence of Barkhausen like jumps in the local magnetic field. These jumps are found in the local magnetization behavior of the sample at temperatures below $T_{c2}$, viz., when the sample is the well inside the FM ordered state. The jumps are associated with the pinning - depinning of the propagating domains as the magnetic field is changed in the ferromagnetic state. These measurements also indicate a coercive field ~ 35 Oe. The jumps become less pronounced above $T_{c2}$ and above $T_{c1}^{*}$ we find (fig.5(d)) that these jumps are absent. The presence of these Barkhausen jumps confirms the presence of long-range FM order present in the sample below $T < T_{c2}$. The long-range FM order results in the formation of relatively large sized domains which can get effectively pinned. Above $T_{c2}$ the weakening of the Barkhausen like jumps suggests that the relatively smaller sized magnetized domains are less effectively pinned compared to that below $T_{c2}$.

The magnetic phase diagram for EuB$_6$ in figs. 4(a) and 4(b), has various boundaries marked as $T_{c2}(H)$ and $T^{*}(H)$. Recall our earlier approximate analysis based on eqn.1 had indicated deviations from a conventional mean-field behavior. We now analyze the bulk magnetization isotherms near the critical temperature in this crystal. It is known that critical scaling exponents near a critical transition temperature are expected to yield information about the nature of magnetic interactions in the system and the universality class associated with the phase transition [34].

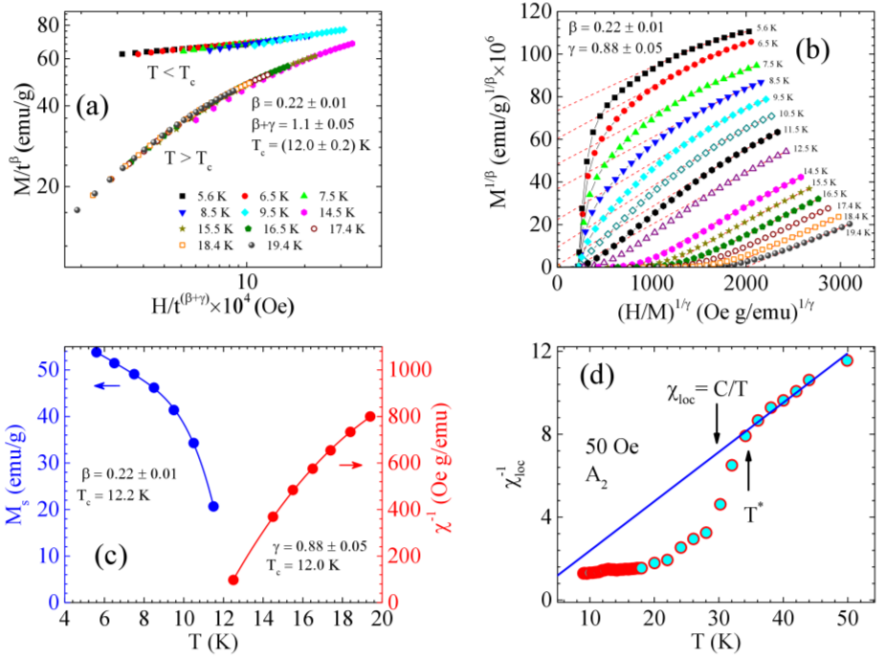

Figure 6: (a) Shows scaling of the M-H curves above and below $T_c$. (b) Modified Arrott plot (MAP) with the critical parameters β and γ determined from the scaling analysis of fig.6(a). (c) Temperature dependencies of $M_s$ (left axis) and $\chi^{-1}$ (right axis) with power-law fitting. See text for details. (d) $\chi_{loc}^{-1}$ versus temperature determined from $<B_z>$ measurements of fig. 3 which were deduced from MOI. $\chi_{loc}^{-1}$ determined from MO is consistent with the values reported in ref.[27].

Around a critical magnetic phase transition, the magnetic equation of state is given as [35], $M(H,t) = |t|^\beta f_\pm(\frac{H}{|t|^{\beta+\gamma}})$, where $t = (T-T_c)/T_c$. In fig.6(a) we show that the isothermal $M(H)$ data captured at different $T$, scales into two distinct curves above and below $T_c$, with the choice of $T_c$ = 12.0 ± 0.2 K, β = 0.22 ± 0.01 and γ = 0.88 ± 0.05 and a $\delta = 1 + \gamma/\beta = 5.0 \pm 0.1$. The values of β, γ and δ differ from the critical exponent values of the universality classes of, the mean field (β = 0.5, γ =1, and δ = 3) and the three dimensional Heisenberg (β = 0.365, γ =1.386, and δ = 4.8) models, while they seem to be close to that for the tricritical mean field model (β = 0.25, γ =1,

and $\delta = 5$) [36,37] though not identical to it. We show in fig.6(b) the modified Arrott's plot (MAP) of $M^{1/\beta}$ versus $(H/M)^{1/\gamma}$ using the values determined from the scaling analysis. With these parameters MAP shows linear parallel lines, thereby supporting the values of $\beta$ and $\gamma$ determined. From backward extrapolation of these parallel lines in MAP, the $T$ dependence of spontaneous magnetization $M_s$ (left axis) and inverse susceptibility $\chi^{-1}$ (right axis) is shown in fig.6(c). In fig.6(c) the intercepts of $M_s(T)$ and $\chi^{-1}(T)$ on the $T$ axis gives an estimate of $T_c = 12.0 \pm 0.2$ K, which is consistent with the scaling analysis of fig.6(a). The temperature dependence of $M_s$ below $T_c$ is governed by the value of $\beta$ and that of $\chi^{-1}$ above $T_c$ is governed by $\gamma$ (whose value is different from 1). The bulk magnetic phase transition temperature $T_c$ which we have determined from these analysis matches closely with $T_{c2}$ determined using our MOI measurements. Here we show another way to determine $T_{c2}$ from the scaling analysis. Below $T_{c2}$ long-range ferromagnetic order is established in EuB$_6$. Note from fig.6(c) that the behavior of $\chi^{-1}$ above $T_c$ isn't completely linear, suggesting the presence of some form of magnetic correlations even above $T_c$. Figure 6(d) shows a plot of the inverse of average local susceptibility ($\chi_{loc}^{-1}$) versus $T$ where the $\chi_{loc}$ is estimated from MOI (see [32] section III in supplementary information). We see from fig.6(d) that the deviation from Curie behavior above $T_c$ is pronounced. We see that although we are above $T_c$, a paramagnetic Curie behavior sets in only at a high temperature $T^*$ (the $T^*$ location is also marked in fig.3(a)-(d)). Figure 6(d) shows that $T^*$ is the temperature above which the sample is in a paramagnetic state as local $<B_z> \to 0$ (see MOI in fig.4(i) and also fig.3). Figure 6(d) shows that below $T^*$ there is a downward curvature of $\chi_{loc}^{-1}$ from the Curie behavior. Some perovskite manganites [38,39,40] have shown deviation from linear $\chi^{-1}$ versus $T$ sustained up to temperatures well above $T_c$ and such systems also show unusual scaling behavior which do not conform to mean field models [38]. Such a behavior in manganite perovskites is similar to the behavior we have shown in EuB$_6$, where, above the critical temperature $T_{c2}$ there is a deviation from Curie behavior until a higher temperature scale $T^*$, and in the critical regime the scaling deviates from mean fields models. In EuB$_6$ the scaling analysis not conforming to mean field models suggests the presence of unusually large critical fluctuations in the system. The presence of these fluctuations is more pronounced above $T_c$ where both bulk and local susceptibilities shows significant deviation from Curie behavior. We believe that the source of larger than expected critical fluctuations in EuB$_6$ are magnetic polaronic clusters created in the system.

We believe that magnetic polarons [18,19,20,21,25] start nucleating in the system below $T^*$. They begin aggregating to form sizeable micron-sized clusters which appear as disjoint magnetized puddles in our MOI (see the yellowish puddles in fig.4(g)). In zero applied field conditions, the clustering of magnetic polaron imparts them with additional stability against thermal fluctuations, due to which fig.6(d) shows a deviation from the Curie behavior below $T^*$. Above T$^*$ there is a paramagnetic Curie behavior of $\chi_{loc}$ (T) (see fig.6(d)). Disorder sites may aid in clustering of these polarons. These stabilized magnetic polaron clusters are the source of the unusually large critical fluctuations discussed earlier, which causes the scaling to deviate from conventional mean field models. With lowering of $T$ below $T^*$ the disjoint magnetized puddles increase in size to reach mesoscopic dimensions wherein they begin to affect electrical transport measurements. The $T_{c1}(H)$ boundary marked in fig.4(b) is a reproduction of the line drawn in refs. [18,20] where a change of slope in the Hall effect is observed and a peak in the nonlinear transport emerges as a result of inhomogeneous current distribution in the sample. The $T_{c1}(H)$ line which is far from the critical phase transition boundary $T_{c2}(H)$, we believe represents a boundary across which polaron clusters while remaining disjoint reach sizeable dimensions spatially, such that they begin to affect the bulk electrical transport properties in the material [18,20, 23]. We believe that it is only below $T_{c1}^*(H)$ which is close to the critical phase transition boundary $T_{c2}(H)$, that the disjoint puddles coalesce to form larger finger-like macroscopic magnetized domains (see MOI images of fig.4). Figure 4(b) shows that with increasing $H$, $T^*(H)$ and $T_{c1}^*(H)$ have opposite slope. It is known from systems like MnSi that magnetic field suppresses critical fluctuations and transforms a zero field first order magnetic transition into a second order one, leading to scaling obeying tricritical mean field model [41,42,43]. The scaling analysis in fig.6, also suggested features associated with the tricritical mean field model for $EuB_6$, which may indicate a field induced modification of the magnetic interactions in this system. This feature is consistent with our earlier approximate study of the critical phenomenon (eqn.1) in fig.3(f) which showed that the critical behavior is affected by magnetic fields (viz., $\alpha$ depends on $H$). We believe that in $EuB_6$ the externally applied magnetic field effectively stabilizes the puddles by suppressing critical fluctuations. Hence at high $H$, each individual magnetized puddle being more stable, not only persist up to higher temperatures, they also remain disjoint without coalescing into larger finger like domains at lower $T$. Due to this the region between $T^*(H)$ and $T_{c1}^*(H)$ in the phase diagrams of figs.4(a) and (b) fans out in area with increasing $H$. Below $T_{c1}^*(H)$ the increased interaction between moments within the large domains, causes local $B_z$ to change slowly with temperature, resulting in the inflection in $<B_z>(T)$ curves (see fig.3 insets). As $T$ is lowered further long-range FM order sets in below $T_{c2}$. The presence of this FM state is confirmed by the observation of the Barkhausen like jumps associated with domain wall pinning, which weaken with increasing temperature as the domains fragment into puddles. We believe that the change in the skewness of the MH curves in fig.1 is related to the domains fragmenting into puddles above $T_{c1}^*$. Thus the transformation into the FM state below $T_{c2}$ occurs in a stepwise manner through nucleation, growth and agglomeration of magnetic polarons at characteristic temperatures.

*Conclusion*

In conclusion, we have imaged the peculiar nature of the onset of a ferromagnetic state in $EuB_6$ through the nucleation and coalescence of local magnetized puddle like regions forming at high temperature. Formation of magnetic polaronic clusters at high temperatures leads to the formation of disjoint magnetized puddles. These puddles lead to deviations from Curie behavior of the temperature dependence of the magnetic susceptibility, above the bulk ferromagnetic ordering temperature. As the bulk ferromagnetic ordering temperature is approached by lowering the temperature, we observe distinct modification in the magnetic domain morphology of the system prior to the onset of large magnetic domains below $T_c$. We believe these changes in domain morphology is due to the agglomeration of magnetic polaronic clusters before the onset of long-range ferromagnetic order below $T_c$. We show via scaling analysis of bulk magnetization data that nature of the universality class suggests magnetic fields may affect critical fluctuations near the magnetic ordering temperature and hence are useful to explain the magnetic field dependence we have shown in the magnetic phase diagram for $EuB_6$. Thus the behavior of magnetic polarons seems to affect the behavior of magnetic domains in this system. The behavior of critical fluctuations in the system, its sensitivity to magnetic fields and the underlying linkages with magnetic polarons is a relevant area for further theoretical and experimental investigation. It may be worthwhile mentioning that magnetic polarons impact the electrical transport properties along with the behavior of magnetic domains, in these strongly correlated electronic systems which have potential for spintronic applications.

*Acknowledgements*

S. S. Banerjee acknowledges the funding support from Department of Science and Technology (TSDP program), Govt. of India and IIT Kanpur.

# Supplementary Information:

# Magneto-optical imaging of stepwise magnetic domain disintegration at characteristic temperatures in EuB$_6$

Dibya J. Sivananda[1], Ankit Kumar[1], Md. Arif Ali[1], S. S. Banerjee[1,*]

[1]*Department of Physics, Indian Institute of Technology Kanpur, Kanpur 208016, India.*

Pintu Das[2]

[2]*Department of Physics, Indian Institute of Technology Delhi, New Delhi 110016, India.*

Jens Müller[3]

[3]*Institute of Physics, Goethe-University Frankfurt, 60438 Frankfurt (M), Germany.*

Zachary Fisk[4]

[4]*Department of Physics, University of California, Irvine, California 92697, USA.*

*email: satyajit@iitk.ac.in

# Section I:

**Sample surface under polarized light:**

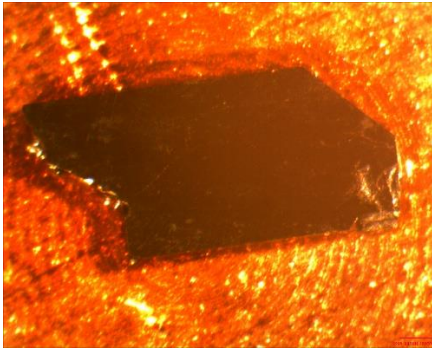

1.5 mm

Figure 1: We show the image sample under polarized light.

We show above an optical image of the sample captured using polarized light. The image shows only the sample topography. We see a fault line in the sample but as mentioned in the text this defect plays no role in the observed results of the paper, and the domain nucleation, propagation, and growth are not correlated with this defect fault line.

# Section II:

**Homogeneity of the sample using EPMA**

EPMA of the $EuB_6$ sample was carried out using Model (manufacturer): JXA-8230(JEOL, USA) at IIT Kanpur. The following table shows the measured percentage of Europium in our $EuB_6$ sample. The variation of the concentrations of Eu over the sample is within a standard deviation of 0.5%. Table 1 shows the Europium percentages for 12 locations within the sample surface. The area's $A_1$ to $A_5$ are shown in Fig. 1(c) of the main manuscript. We reproduce that Fig. 1(c) below for convenience.

| Element | Eu Percentage |
|---|---|
| *Europium (Area 1)* | *70.100* |
| | *69.124* |
| | *69.613* |
| *Europium (Area 2)* | *69.555* |
| | *69.843* |
| | *69.649* |
| *Europium (Area 3)* | *68.485* |
| | *69.667* |
| | *70.111* |
| *Europium (Area 5)* | *69.648* |
| | *69.802* |
| | *68.769* |

Thus, we find that the chemical stoichiometry across the sample is homogeneous.

Table 1: Shows the Eu concentrations at twelve locations distributed inside the area $A_1$, $A_2$, $A_3$ and $A_5$.

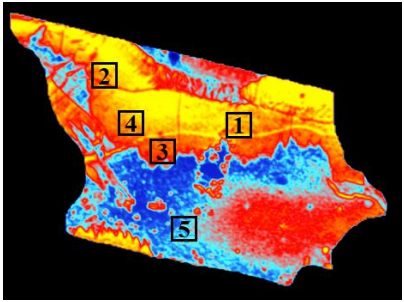

Fig.2: MOI image showing area $A_1$, $A_2$, $A_3$, $A_4$ and $A_5$.

# Section III:

**Magneto-Optic Imaging**

In the presence of a magnetic field when a plane polarized light traverse through certain media the plane of polarization of light rotates, and this effect is called Faraday Rotation. Linearly polarized light consists of the superposition of right circularly polarized light and left circularly polarized light, both of which shifts in phase as the light ray travels through certain media in presence of magnetic fields. This happens because left and right circularly polarized light propagate with different velocities inside the medium in the presence of a magnetic field. Upon exiting the medium, the two left and right circularly polarized light superimpose again to form a plane polarized light but now with the plane of polarization rotated by an angle θ. Where θ is given by:

$$\theta = V B_z d \quad \quad \quad (1)$$

Where $V$ is the Verdet's constant, $B_z$ is the component of the magnetic field along the direction of the light and $d$ is the distance which the light wave traverses in the medium. This relation tells us that the rotation of the plane of polarization of light depends on the local magnetic field inside the medium. So, by measuring the angle of rotation of the plane of polarization, we can determine the local magnetic fields inside the medium. We use this effect in our MOI setup to measure the local magnetic fields on a sample placed in a magnetic field.

Our samples under investigation do not possess high Faraday Optic Effect, hence we put some other material with high Verdet's constant on our sample to enhance the effect. We call such material magneto-optic indicator. We use ferrimagnetic Bismuth doped iron garnet films in our magneto-optic indicator which has a very high Verdet's constant. In our experiments, we have used Bi-doped yttrium iron garnet film (YIG) grown on [100] oriented gadolinium-gallium-garnet (GGG) substrate. We use a 0.5 mm thick GGG substrate to grow the YIG films. Below we show the cross-sectional schematic of our multilayered MO active indicator film.

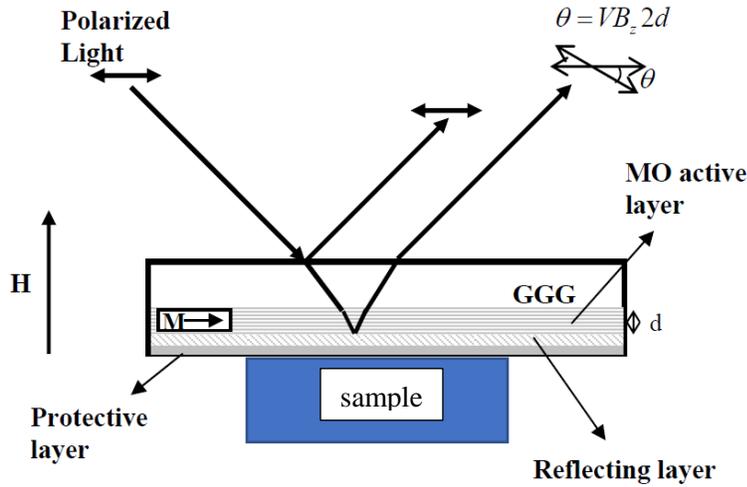

Figure 3: Schematic of the magneto-optic indicator

An aluminum layer of a few hundred nanometers is coated on top of the YIG layer to act as the reflector which is again protected from scratches by evaporating a thin layer of titanium dioxide. The typical thickness of the magneto-optic active layer (YIG layer) is ~ 3-6 μm. While imaging one has to ensure that the sample surface being imaged has to be as smooth as possible so that the distance between the magneto-optic indicator and the sample surface is minimized. The freshly cleaved $EuB_6$ samples we use for our study have a surface roughness of less than 0.1 microns. The optimum distance for highest sensitivity is achieved when the separation between the magneto-optic active layer (YIG) and the sample surface is within 1-3 μm. The saturation field of the magneto-optic active layer is of the order of a few thousand Gauss, and hence eqn.1 holds for any field below that. The Verdet's constant for our magneto-optic active layer is maximum for a wavelength of 500 nm. The YIG films used for our measurement has a Verdets constant of ~ $9.05 \times 10^{-4}$ deg.Oe$^{-1}$μm$^{-1}$. As the Verdets constants is wavelength sensitive, its value being maximum for green wavelength, therefore for our experiments we work with 550nm ±10nm green light. We have an effective lateral spatial resolution of 0.7 μm in our setup.

We show below the basic schematic of our magneto-optic set up. The set up mainly comprises a Carl-Zeiss Axiotech vario polarized light microscope and a Charged Coupled Device (CCD from Andor, Model: Andor iXON) camera. A 100W halogen lamp emits white light which passes through a bandpass green filter set at 550 nm. The monochromatic green light passes through a linear polarizer (P), and is then incident on a beam splitter (B) which reflects the light onto the MO indicator (V) which is placed directly over our sample. The polarized light undergoes faraday rotation after passing through the MO-active layer. It is then reflected back by the aluminum mirror through an analyzer (A) which is kept in a cross position (perpendicular) with respect to the polarizer (P). The light coming out of the analyzer is then incident on a high sensitive CCD camera with a Quantum efficiency of > 90% at 550 nm (Andor Ixon). Our CCD camera contains a silicon-based semiconductor chip having a two-dimensional matrix of $512 \times 512$ photo-sensors or pixels with the area of one pixel being $16 \mu m^2$. We use a 14-bit resolution camera, i.e., it has $2^{14}$ (16384) gray scale values. The intensity captured by the CCD camera after the light passes through the polarizer analyzer combination is given by Malus law, taking into account the light absorbed by the MO layer is given by:

$$I = I_0 \exp(-2 \gamma d) \sin^2 \theta \quad \ldots\ldots\ldots\ldots\ldots\ldots\ldots\ldots\ldots \quad (2)$$

where $\gamma$ is the absorption coefficient of the MO layer, and $\theta$ is as expressed above. In general, $\theta$ is very small, and hence eqn.2 can be written as:

$$I \sim I_c \theta^2. \quad \ldots\ldots\ldots\ldots\ldots\ldots\ldots\ldots\ldots\ldots \quad (3)$$

where $I_c = I_o \exp(-2\gamma d)$.

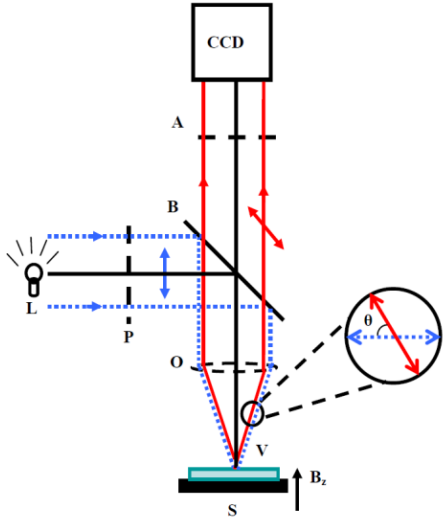

Figure 4: Ray diagram schematic of MO imaging setup. It consists of light source (L), a linear polarizer (P), a beam splitter (B), objective (O), MO indicator (V), sample(S), and analyzer (A). Also indicated is the local magnetic field direction ($B_z$). The blow-up shows the state of linear polarization in the incident beam (dotted line) and reflected beam (solid line).

Substituting, eqn.1, $\theta = V B_z d$, in eqn.3 gives

$I \sim I_c (Vd)^2 B_z^2$

Using the above equation and by applying calibrated $B$ values, from the intensity distribution $I(x,y)$ the local $B_z(x,y)$ information can be obtained.

Determining local susceptibility using MOI.

The above shows that MOI imaging gives information about the local magnetic field $B$ in a sample. Equivalently the MOI can also be used to determine the average local susceptibility in a particular region of the sample. Since in MOI, the faraday rotated light intensity in a particular region of the sample is proportional to the square of the local field and since the local field $B = (1 + \chi_{loc})H$, where $H$ is applied field and $\chi_{loc}$ is the average local susceptibility, therefore $I \sim I'_c(1 + \chi_{loc})^2 H^2$. Since the value of the applied field, $H$ is known and by calibrating the magneto-optic intensities at high temperatures > 100 K where the sample is in the paramagnetic state using susceptibility values determined from bulk magnetization measurements, we determine the average local susceptibility using MOI. The average local susceptibility $\chi_{loc}$ is used to denote the local susceptibility determined by the above procedure and averaging the value over a $50 \times 50$ μm² area in the MOI image.

## Section IV:

### Phase diagrams for A₃, A₄ and A₅:

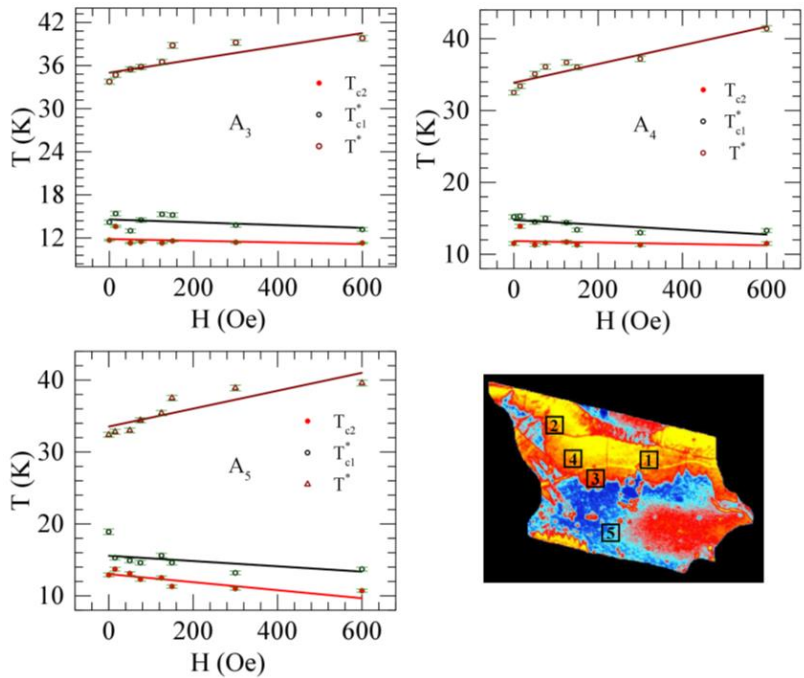

Fig.5: Phase diagrams of area A₃, A₄ and A₅.

Similar to the phase diagrams for area's $A_1$ and $A_2$ (shown in the paper in Figs. 4(a) and 4(b)), above we have shown the phase diagrams for area $A_3$, $A_4$, and $A_5$ in-order to show that different parts of the sample have similar behavior. We see approximately identical behavior all over the sample surface which is possible only if the sample has a high degree of homogeneity in chemical stoichiometry. We see that area $A_3$, $A_4$ and $A_5$ also have an increasing slope of $T^*$ as shown in area $A_1$ and $A_2$ in the main text of the paper. Moreover, $T_{c1}^*$ and $T_{c2}$ have a decreasing slope for area $A_3$, $A_4$, and $A_5$. We see that in all the different regions of the sample the values of the characteristic temperatures lie within the values of the error bars.

## Section V:

**Alternative criteria to determine $T_{c2}$:**

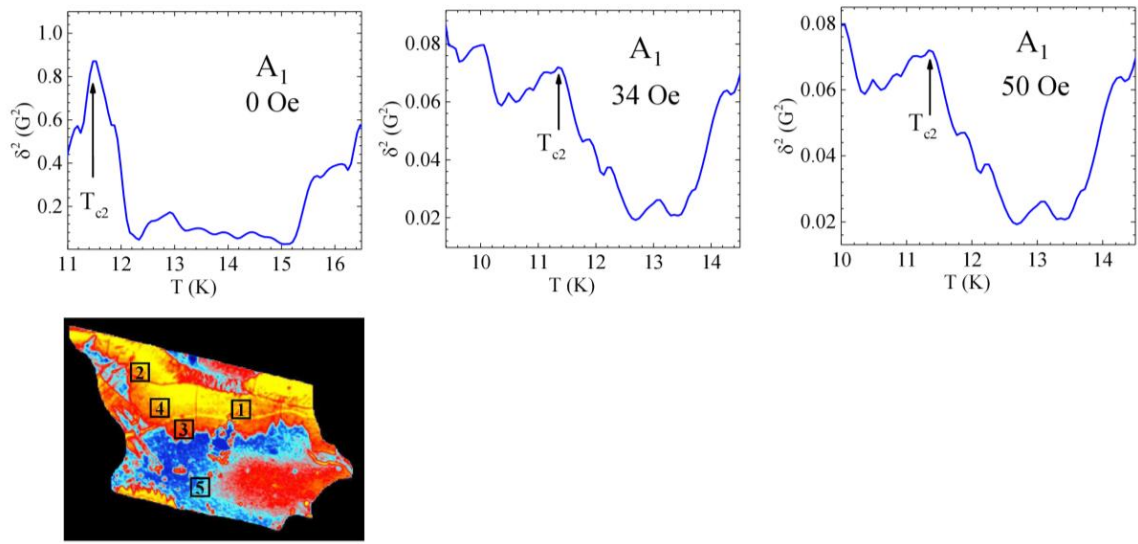

Figure 6: Shows the $\delta^2(G^2)$ v/s T curves for $A_1$ at different fields.

Here we show an alternative criteria to determine $T_{c2}$. Figure 6 shows the $\delta^2(G^2)$ v/s T curves for area $A_1$ at different fields. We observe a sudden increase in $\delta^2(G)$ when we cross the two critical temperatures $T_{c1}$ and $T_{c2}$. Here we show that we can determine $T_{c2}$ from the position in the $\delta^2(G^2)$ v/s T curve at which $\delta^2(G^2)$ increases. Beyond $T_{c2}$ as we increase the temperature as seen from Fig. 6, $\delta^2(G^2)$ is suppressed. In between $T_{c1}^*$ and $T_{c2}$ the $\delta^2(G^2)$ is suppressed and increases as we approach $T_{c1}^*$.